\documentclass[aps,prb,twocolumn,superscriptaddress,floatfix]{revtex4}
\usepackage{color}
\usepackage{graphicx}
\usepackage{epsfig}
\usepackage{textcmds}
\bibstyle{apsrev.bib}

\begin{document}
\input epsf.sty

\title{Rounding of first-order phase transitions and optimal cooperation in scale-free networks}

\author{M. Karsai}
\affiliation{
Institute of Theoretical Physics,
Szeged University, H-6720 Szeged, Hungary}
\affiliation{
Centre de Recherches sur les Tr\'es Basses
Temp\'eratures\thanks{U.P.R. 5001 du CNRS, Laboratoire conventionn\'e
avec l'Universit\'e Joseph Fourier}, B. P. 166, F-38042 Grenoble,
France}
\author{J-Ch. Angl\`es d'Auriac}
\affiliation{
Centre de Recherches sur les Tr\'es Basses
Temp\'eratures\thanks{U.P.R. 5001 du CNRS, Laboratoire conventionn\'e
avec l'Universit\'e Joseph Fourier}, B. P. 166, F-38042 Grenoble,
France}
\author{F. Igl\'oi}
\affiliation{
Research Institute for Solid State Physics and Optics,
H-1525 Budapest, P.O.Box 49, Hungary}
\affiliation{
Institute of Theoretical Physics,
Szeged University, H-6720 Szeged, Hungary}

\date{\today}

\begin{abstract}
We consider the ferromagnetic large-$q$ state Potts model in complex evolving networks,
which is equivalent to an optimal cooperation problem, in which the agents try to
optimize the total sum of pair cooperation benefits and the supports of independent projects.
The agents are found to be typically of two kinds: a fraction of $m$ (being the
magnetization of the Potts model) belongs to a large
cooperating cluster, whereas the others are isolated one man's projects.
It is shown rigorously that the homogeneous model has a strongly first-order phase transition,
which turns to second-order for random interactions (benefits), the properties
of which are studied numerically on the Barab\'asi-Albert network. The distribution of finite-size
transition points is characterized by a shift exponent, $1/\tilde{\nu}'=.26(1)$, and by a
different width exponent, $1/\nu'=.18(1)$, whereas the magnetization at the transition point
scales with the size of the network, $N$, as: $m\sim N^{-x}$, with $x=.66(1)$.
\end{abstract}

\maketitle

\newcommand{\bc}{\begin{center}}
\newcommand{\ec}{\end{center}}
\newcommand{\be}{\begin{equation}}
\newcommand{\ee}{\end{equation}}
\newcommand{\beqn}{\begin{eqnarray}}
\newcommand{\eeqn}{\end{eqnarray}}

\section{Introduction}

Complex networks have been used to describe the structure and topology of
a large class of systems in different
fields of science, technics, transport, social and political life,
etc, see Refs.\cite{AB01,DM01,DM03,N03} for recent reviews.
A complex network is represented by a graph\cite{bollobas}, in which
the nodes stand for the agents and the edges denote the possible
interactions. Realistic networks generally have three basic properties.
The average distance between the nodes is small, which is called the small-world effect\cite{WS98}.
There is a tendency of clustering and the degree-distribution of the edges, $P(k)$,
has a power-law tail, $P_D(k) \simeq A k^{-\gamma},~ k \gg 1$.
Thus the edge distribution is scale free\cite{BA99}, which is usually attributed to
growth and preferential attachment during the evaluation of the network.

In reality there is some sort of interaction between the agents of a network
which leads to some kind of cooperative behavior in macroscopic scales. 
One throughly studied question in this field is the spread of infections and epidemics
in networks\cite{epidemic,newman,dezso}, which problem is closely related to another non-equilibrium processes,
such as percolation\cite{percolation}, diffusion\cite{diffusion}, the contact process\cite{kji06} or the zero-range process\cite{zrp}, etc.
In another investigations one considers simple magnetic models\cite{stauffer,ising,sajat,potts,joseph},
in which the agents are represented by classical (Ising or Potts) spin variables,
the interactions are described by ferromagnetic couplings, whereas the temperature plays
the r\^ole of a disordering field.

In theoretical investigations of the cooperative behavior one usually resort on
some kind of approximations. For example the sites of
the networks are often considered uncorrelated, which is generally not true for evolving
networks, such as the Barab\'asi-Albert (BA) network. However this effect is expected to be
irrelevant, as far as the singularities in the system are considered. Also the simple
mean-field approach could lead to exact results due to long-range interactions in the
networks, which has been checked by numerical simulations\cite{stauffer} and by another, more accurate
theoretical methods\cite{joseph} (Bethe-lattice approach, replica method, etc.). In these calculations
the critical behavior of the network is found to depend on the value of the degree exponent,
$\gamma$. For sufficiently weakly connected networks with $\gamma>\gamma_u$ ($\gamma_u=5$
for the Ising model) there are conventional mean-field singularities. In the
intermediate or unconventional mean-field regime, for $\gamma_u>\gamma>\gamma_c$, the critical exponents
are $\gamma$ dependent. Finally, for $\gamma
\le \gamma_c$, when the average of $k^2$, defined by $\langle k^2 \rangle=\int P_D(k) k^2 {\rm d} k$, as
well as the strength of the average interaction becomes divergent the scale-free
network remains in the ordered state at any finite temperature. Since $\gamma_c=3$,
in realistic networks with homogeneous interactions always this type of phenomena is expected to occur.
In weighted networks, however, in which the strengths of the interaction is appropriately
rescaled with the degrees of the connected vertices, $\gamma_c$ is shifted to larger values
and therefore the complete phase-transition scenario can be tested\cite{joseph,kji06}.
We note that the properties of the phase transitions are generally different for undirected
(as we consider here) and directed networks\cite{sumour}.

In several models the phase transition in regular lattices is of first order, such
as for the $q$-state Potts model for sufficiently large value of $q$. Putting these models on
a complex network the inhomogeneities of the lattice play the r\^ole of some kind of disorder
and it is expected that the value of the latent heat is reduced or even the transition is
smoothened to a continuous one. This type of scenario is indeed found in a mean-field treatment\cite{sajat},
in which the transition is of first-order for $\gamma>\gamma(q)$ and becomes continuous
for $\gamma_c<\gamma<\gamma(q)$, where $3<\gamma(q)<4$. On the other hand in an effective medium
Bethe lattice approach one has obtained $\gamma(q)=3$, thus the unconventional mean-field regime is
absent in this treatment\cite{potts}.

The interactions considered so far were homogeneous, however, in realistic situations
the disorder is inevitable, which has a strong influence on the properties of the
phase transition. In regular lattices and for a second-order transition Harris-type relevance-irrelevance criterion can be used to decide about the stability of the pure system's fixed point in the presence
of weak disorder. On the contrary for a first-order transition such type of criterion does
not exist. In this case rigorous results asserts that in two
dimensions (2d) for any type of continuous disorder the originally first order
transition softens into a second order one~\cite{aizenmanwehr}. In three dimensions there are
numerical investigations which have
shown\cite{uzelac,pottssite,pottsbond,mai05,mai06} that this kind of softening takes place only for
sufficiently strong disorder. 

In this paper we consider interacting models with random interactions on complex networks
and in this way we study the combined effect of network topology and bond disorder. The
particular model we consider is the random bond ferromagnetic Potts model (RBPM)
for large value of $q$. This model besides its relevance in ordering-disordering phenomena
and phase transitions has an exact relation with an optimal cooperation problem\cite{aips02}.
This mapping is based on the observation that in the large-$q$ limit the thermodynamic
properties of the system are dominated by one single diagram~\cite{JRI01} of the high-temperature expansion~\cite{kasteleyn} and its calculation is equivalent to the solution of an optimization
problem. This optimization problem can be interpreted
in terms of cooperating agents which try to maximize the total sum of benefits received for pair
cooperations plus a unit support which is paid for each independent projects. For a given realization of the
interactions the optimal state is calculated exactly by a combinatorial optimization algorithm
which works in strongly polynomial time\cite{aips02}.
The optimal graph of this problem consists of connected components (representing sets of cooperating
agents) and isolated sites and its temperature (support) dependent topology
contains information about the collective behavior of the agents. In the thermodynamic
limit one expects to have a sharp phase transition in the system, which separates the ordered
(cooperating) phase with a giant clusters from a disordered (non-cooperating) phase, having
only clusters of finite extent.

The structure of the paper is the following. The model and the
optimization method used in the study for large $q$ is presented in
Sec.~\ref{sec_model}. The solution for homogeneous non-random evolving networks
can be found in Sec.~\ref{sec_hom}, whereas numerical study of the random model on the
Barab\'asi-Albert network is presented in Sec.~\ref{sec_numerics}. Our results are discussed
in Sec.~\ref{sec_disc}.

\section{The model and its relation with optimal cooperation}
\label{sec_model}

The $q$-state Potts model~\cite{Wu} is defined by the Hamiltonian:
\begin{equation}
\mathcal{H}=-\sum_{\left\langle i,j\right\rangle }J_{ij}\delta(\sigma_{i},\sigma_{j})
\label{eq:hamilton}
\end{equation}
in terms of the Potts-spin variables, $\sigma_{i}=0,1,\cdots,q-1$. Here
$i$ and $j$ are sites of a lattice, which is represented by a complex network in our case
and the summation runs over nearest neighbors, i.e. pairs of connected sites.

The couplings, $J_{ij}>0$, are ferromagnetic and they are either identical, $J_{ij}=J$,
which is the case of homogeneous networks, or they are identically and independently distributed
random variables. In this paper we use a quasi-continuous  distribution:
\begin{equation}
P(J_{ij})= \frac{1}{l} \sum_{i=1}^l \delta\left[J\left(1+\Delta\frac{2i-l-1}{2l}\right)-J_{ij}\right] \label{eq:distr}
\end{equation}
which consists of large number of $l$ equally spaced discrete values within the range
$J(1 \pm \Delta/2)$ and $0 \le \Delta \le 2$ measures the strength of disorder.

For a given set of couplings the partition function of the system is convenient to write in
the random cluster representation~\cite{kasteleyn} as:
\begin{equation}
Z =\sum_{G}q^{c(G)}\prod_{ij\in G}\left[q^{\beta J_{ij}}-1\right]
\label{eq:kasfor}
\end{equation}
where the sum runs over all subset of bonds, $G$ and $c(G)$ stands for
the number of connected components of $G$. In Eq.~(\ref{eq:kasfor}) we use
the reduced temperature, $T \to T \ln q$ and its inverse $\beta
\to \beta/\ln q$, which are of $O(1)$ even in the large-$q$ limit~\cite{long2d}. In this limit we have
$q^{\beta J_{ij}} \gg 1$ and the partition function can be written as
\begin{equation}
Z=\sum_{G\subseteq E}q^{\phi(G)},\quad \phi(G)=c(G) + \beta\sum_{ij\in G} J_{ij}\label{eq:kasfor1}
\end{equation}
which is dominated by the largest term, $\phi^*=\max_G \phi(G)$. 
Note that this graph, which is called the optimal set,
generally depends on the temperature. The
free-energy per site is proportional to $\phi^*$ and given by $-\beta
f= \phi^*/N$ where $N$ stands for the number of sites of the lattice.

As already mentioned in the introduction the optimization in Eq.~(\ref{eq:kasfor1}) can be interpreted as an
optimal cooperation problem~\cite{aips02} in which the agents, which
cooperate with each other in some projects, form connected components. Each cooperating
pair receives a benefit represented by the weight of the connecting edge (which is
proportional to the inverse temperature) and also
there is a unit support to each component, i.e. for each projects. Thus by uniting two
projects the support will be reduced but at the same time the edge benefits will be enhanced.
Finally one is interested in the optimal form of cooperation when the total value of the project grants
is maximal.

In a mathematical point of view the
cost-function in Eq.~(\ref{eq:kasfor1}), $-\phi(G)$, is sub-modular~\cite{gls81} and there is an
efficient combinatorial optimization algorithm which calculates the optimal set
(i.e. set of bonds which minimizes the cost-function)
exactly at any temperature in strongly polynomial time~\cite{aips02}. In the algorithm the
optimal set is calculated iteratively and at each step one new vertex of the lattice is taken
into account. Having the optimal set at a given step, say with $n$ vertices, its connected
components have the property to contain all the edges between their sites. Due to the submodularity
of $-\phi(G)$ each
connected component is contracted into a new vertex with effective weights being the sum
of individual weights in the original representation. Now adding a new vertex one should
solve the optimization problem in terms of the effective vertices, which needs the application
of a standard maximum flow algorithm, since any contractions should include the new vertex.
After making the possible new contractions one repeats
the previous steps until all the vertices are taken into account and the optimal set of the
problem is found.

This method has already been applied for 2d\cite{ai03,long2d} and 3d\cite{mai05,mai06} regular
lattices with short range random interactions. As a general result
the optimal graph at low temperatures is compact and the
largest connected subgraph contains a finite fraction of the sites, $m(T)$,
which is identified by the orderparameter of the system. In
the other limit, for high temperature, most of the sites in the
optimal set are isolated and the connected clusters have a finite
extent, the typical size of which is used to define the correlation
length, $\xi$. Between the two phases there is a sharp phase
transition in the thermodynamic limit, the order of which depends on the dimension of
the lattice and the strength of disorder, $\Delta$.

In the following the optimization problem is solved exactly for homogeneous evolving networks
in Sec.\ref{sec_hom} and studied numerically in random Barab\'asi-Albert networks in Sec.\ref{sec_numerics}.

\section{Exact solution for homogeneous evolving networks}
\label{sec_hom}

In regular $d$-dimensional lattices the solution of the
optimization problem in Eq.~(\ref{eq:kasfor1}) is simple, since there are only
two distinct optimal sets, which correspond to the $T=0$ and $T \to \infty$
solutions, respectively. For $T<T_c(0)$ it is the
fully connected diagram, $E$, with a free-energy: $-\beta N f=1+N\beta J d$
and for $T>T_c(0)$ it is the empty diagram, $\O$, with $-\beta N f=N$. In the
proof we make use of the fact, that any edge of a regular lattice, $e_1$, can be
transformed to any another edge, $e_2$, through operations of
the automorphy group of the lattice. Thus if $e_1$ belongs to some optimal set, then
$e_2$ belongs to an optimal set, too. Furthermore, due to submodularity the union of
optimal sets is also an optimal set, from which follows that at any temperature the optimal
set is either $\O$ or $E$.
By equating the free energies in the two phases we obtain for the position of the transition point: $T_c(0)=Jd/(1-1/N)$
whereas the latent heat is maximal: $\Delta e/T_c(0)=1-1/N$.

In the following we consider the optimization problem in homogeneous evolving networks which are generated by the following rules:
\begin{itemize}
\item
we start with a complete graph with $2\mu$ vertices 
\item
at each timestep we add a new vertex
\item
which is connected to $\mu$ existing vertices.
\end{itemize}
In definition of these networks there is no restriction in which way
the $\mu$ existing vertices are selected. These
could be chosen randomly, as in the Erd\H os-R\'enyi model\cite{ER},
or one can follow some defined rule, like the preferential
attachment in the BA network\cite{BA99}. In the following we show that for such networks the
phase-transition point is located at $T_c(0)=J\mu$ and
for $T<T_c(0)$ ($T>T_c(0)$) the optimal set is the fully connected diagram (empty diagram), as
for the regular lattices. Furthermore, the latent heat is maximal: $\Delta e/T_c(0)=1$.

In the proof we follow the optimal cooperation algorithm\cite{aips02} outlined in Sec.\ref{sec_model},
and in application of the algorithm we add the vertices one by one
in the same order as in the construction of the network.
First we note, that the statement is true for the initial graph, which is a complete graph, thus the
optimal set can be either fully connected, having a free-energy: $-\beta 2\mu f=1+\mu(2\mu-1)\beta J d$,
or empty, having a free-energy: $-\beta 2\mu f=2\mu$, thus the transition point is indeed at $T=T_c(0)$. We
suppose then that the property is satisfied after $n$ steps and add a new vertex, $v_0$. Here we
investigate the two cases, $T \le T_c(0)$ and $T \ge T_c(0)$ separately.

\begin{itemize}

\item
If $T \le T_c(0)$, then according to our statement all vertices of the original graph are contracted
into a single vertex, $s$, which has an effective weight, $\mu \times J/T > \mu J/T_c(0)=1$, to the new vertex $v_0$. Consequently in the optimal set $s$ and $v_0$ are connected, in accordance with our statement.

\item
If $T \ge T_c(0)$, then all vertices of the original graph are disconnected, which means that
for any subset, $S$, having $n_s \le n$ vertices and $e_s$ edges one has: $n_s \ge e_s JT +1$.
Let us denote by $\mu_s \le \mu$ the number of edges between $v_0$ and the vertices of $S$. One
has $\mu_s J/T \le \mu J/T \le \mu J/T_c(0)=1$, so that for the composite $S+v_0$ we have:
$n_s+1 \ge e_s JT +1 + \mu_s J/T$, which proves that the vertex $v_0$ will not be connected
to any subset $S$ and thus will not be contracted to any vertex.

\end{itemize}

This result, i.e. a maximally first-order transition of the large-$q$ state Potts model holds for
a wide class of evolving networks, which satisfy the construction rules presented above. This is true,
among others, for randomly selected sites, for the BA evolving network which has a degree exponent $\gamma=3$
and for several generalizations of the BA network\cite{AB01} including nonlinear preferential attachment,
initial attractiveness, etc. In these latter network models the degree exponent can vary in a range
of $2<\gamma<\infty$. It is interesting to note that for uncorrelated random networks with a given
degree distribution the $q$-state Potts model is in the ordered phase\cite{sajat,potts} for any $\gamma \le 3$.
This is in contrast to evolving networks in which correlations in the network sites results in the existence
of a disordered phase for $T>T_c(0)$, at least for large $q$.

\section{Numerical study of random Barab\'asi-Albert networks}
\label{sec_numerics}

In this section we study the large-$q$ state Potts model in the BA network with a given
value of the connectivity, $\mu=2$, and the size of the network varies between $N=2^6$ to $N=2^{12}$.
The interactions are independent random variables taken from the quasi-continuous distribution
in Eq.(\ref{eq:distr}) having $l=1024$ discrete peaks and we fix $J=1$. The advantage of using
quasi-continuous distributions is that in this way we avoid
extra, non-physical singularities, which could appear for discrete (e.q. bimodal) distributions\cite{long2d}.
For a given size we have generated
$100$ independent networks and for each we have $100$ independent realizations of the disordered
couplings.

\subsection{Magnetization and structure of the optimal set}
\begin{figure}
  \begin{center}
     \includegraphics[width=3.35in,angle=0]{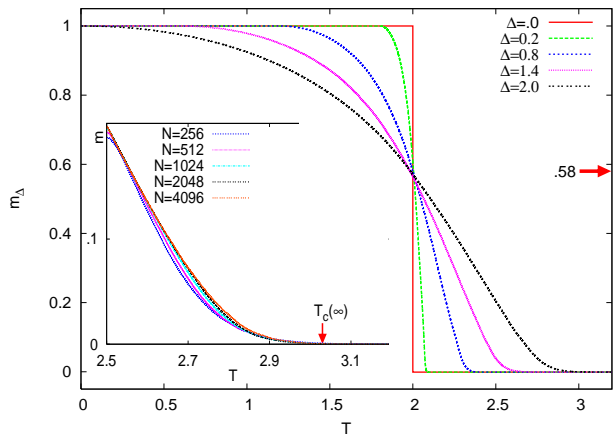}
   \end{center}
   \caption{(Color online) Temperature dependence of the average magnetization
in a BA network of $N=1024$ sites for different strength of the disorder, $\Delta$. At
$T=T_c(0)=2$ the magnetization is independent of $\Delta>0$ and its value is indicated by
an arrow. Inset: The average magnetization for uniform disorder, $\Delta=2$, close to the transition point
for different finite sizes. The arrow indicates the critical point of the infinite system.
}
   \label{magn}
 \end{figure}

In Fig.\ref{magn} the temperature dependence of the average magnetization is shown for various strength of
disorder, $\Delta$, for a BA network of $N=1024$ sites. It is seen that the sharp first-order phase
transition of the pure system with $\Delta=0$ is rounded and the magnetization has considerable variation
within a temperature range of $\sim \Delta$. The phase transition seems to be continuous even
for weak disorder. Close to the transition point the magnetization curves for uniform disorder ($\Delta=2$)
are presented in the inset of Fig.\ref{magn}, which are calculated for different finite systems.

\begin{figure}
  \begin{center}
     \includegraphics[width=3.in,angle=0]{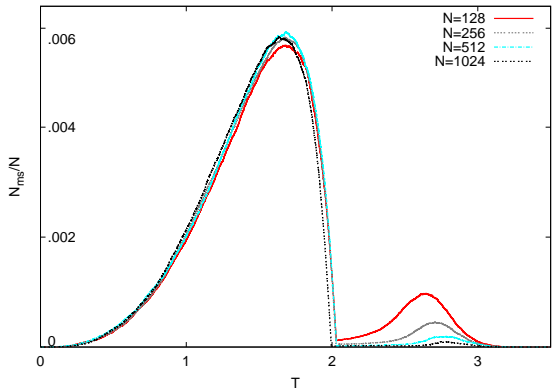}
   \end{center}
   \caption{(Color online) Fraction of intermediate size clusters as a function
of the temperature.
}
   \label{small_cl}
 \end{figure}

Some features of the magnetization curves and the properties of the phase transition
can be understood by analysing the structure of the
optimal set. For low enough temperature this optimal set is fully connected, i.e. the magnetization is $m=1$,
which happens for $T<T_c(0)-\Delta$.
Indeed, the first sites with $k=\mu=2$ (i.e. those which have only outgoing edges)
are removed from the fully connected diagram, if the sum of the
connected bonds is $\sum_{i=1}^{\mu} J_i < T$, which happens within the temperature range indicated
above. From a similar analysis follows that the optimal set is empty for any finite system
for $T>T_c(0)+\Delta$. The magnetization can be estimated for
$t=T-(T_c(0)-\Delta) \ll 1$, and the correction is given by: $1-m \sim t^{\mu}$. For the numerically
studied model with $\mu=2$ and $\Delta=2$, we have $m(T) \approx 1-T^2/8$, which is indeed a good
approximation for $T<1$. In the temperature range $T_c(0)-\Delta<T<T_c(0)+\Delta$ typically the sites
are either isolated or belong to the largest cluster. There are also some clusters with an
intermediate size, which are dominantly two-site clusters for $T<T_c(0)$ and their fraction is less
then $1\%$, as
shown in Fig.\ref{small_cl}. The fraction of two-site clusters for $\Delta=2$ and $T<T_c(0)=2$
can be estimated as follows. First, we note that since they are not part of the biggest
cluster they can be taken out of a fraction of $p_1 = 1-m(T)$ sites. Before being disconnected
a two-site cluster has typically three bonds to the biggest cluster, denoted by $J_1$,
$J_2$ and $J_3$. When it becomes disconnected we have $J_1+J_2+J_3<T$, which happens with a
probability $p_2=T^3/48$. At the same time the coupling within the two-site cluster should be $J_4>T$,
which happens with probability $p_3=(2-T)/2$. Thus the fraction of two-site clusters is approximately:
$n_2 \approx p_1\times p_2 \times p_3 \approx T^5 (2-T)/768$, which describes well the general
behavior of the distribution in Fig.\ref{small_cl}.

\begin{figure}
  \begin{center}
     \includegraphics[width=3.35in,angle=0]{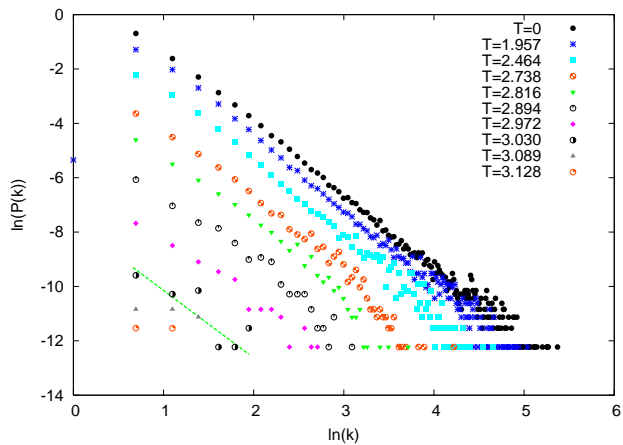}
   \end{center}
   \caption{(Color online) Degree distribution of the largest cluster at different temperatures
in a finite network with $N=2048$.
The dashed straight line indicates the range of the critical temperature.}
   \label{degree}
 \end{figure}

In the temperature range $T>T_c(0)$ the intermediate clusters have at least three sites and their
fraction is negligible, which is seen in Fig.\ref{small_cl}.
Consequently the intermediate size clusters do not influence the properties of the phase
transition in the system.
In the ordered phase, $T<T_c$, the largest connected cluster contains a finite fraction of 
$m(T)<1$ of the sites. We have analyzed the degree distribution of this connected giant
cluster in Fig. \ref{degree}, which has scale-free behavior and for any temperature $T<T_c$ there is the
same degree exponent, $\gamma=3$, as for the original BA network.
We note an interesting feature of the magnetization curves in  Fig.\ref{magn}
that cross each other at the transition
point of the pure system, at $T_c(0)=2$, having a value of $m(T_c(0))=0.58$, for any strength of
disorder. This property follows from the fact that for
a given realization of the disorder the optimal set at $T=T_c(0)$ only depends on the sign
of the sum of fluctuations of given couplings (c.f. some set of sites is connected (disconnected)
to the giant cluster only for positive (negative)
accumulated fluctuations) and does not depend on the actual value of $\Delta>0$.

We can thus conclude the following picture about the evaluation of the optimal set.
This is basically one large connected cluster with ${\cal N}$ sites, immersed in the see of
isolated vertices. With increasing temperature more and more loosely connected sites are
dissolved from the cluster, but for $T<T_c$ we have ${\cal N}/N=m(T)>0$ and the cluster
has the same type of scale-free character as the underlaying network. On the contrary above the phase-transition point, $T_c(0)+\Delta>T>T_c$, the large cluster has only a finite
extent, ${\cal N} < \infty$. The order of the transition depends on the way how ${\cal N}$ behaves
close to $T_c$. A first-order transition, i.e. phase-cooexistence at $T_c$ does not fit to
the above scenario. Indeed, as long as ${\cal N} \sim N$ the same type of continuous erosion
of the large cluster should take place, i.e. the transition is of second order for any strength,
$\Delta>0$ of the disorder. Approaching the critical point one expects the following
singularities: $m(T) \sim (T_c-T)^{\beta}$ and ${\cal N} \sim (T-T_c)^{-\nu'}$. Finally, at $T=T_c$
the large cluster has ${\cal N} \sim N^{1-x}$ sites, with $x=\beta/\nu'$.

\subsection{Distribution of the finite-size transition temperatures}
\label{sec_distr}

The first step in the study of the critical singularities is to locate the position of the phase-transition
point. In this respect it is not convenient to use the magnetization, which approaches zero
very smoothly, see the inset of Fig.\ref{magn}, so that there is a relatively large error by calculating $T_c$
in this way. One might have, however, a better estimate by defining for each given sample, say $\alpha$,
a finite-size transition temperature $T_c(\alpha,N)$, as has been made for regular
lattices\cite{long2d,mai05,mai06}. For a network we use a condition for the size of the
connected component: ${\cal N}(T) \simeq A N^{1-x}$ , in which $x$ is the magnetization critical exponent and $A=O(1)$ is a free parameter, from which the scaling form of the distribution is expected to be independent.
The calculation is made self-consistently: for a fixed $A$ and
a starting value of $x_s=x_1$ we
have determined the distribution of the finite-size transition temperatures and at
their average value we have obtained an estimate for the exponent, $x=x_2$.
Then the whole procedure is repeated with $x_s=x_2$, etc. until a
good convergence is obtained. Fortunatelly the distribution function, $p(T_c,N)$, has only a weak
$x$-dependence thus it was enough to make only two iterations. We have started with a logarithmic initial condition, ${\cal N}(T) \simeq A \ln N$,
which means formally $x_1=1$ and we have obtained $x_2=.69$. Then in the
next step the critical exponents are converged within the error
of the calculation and they are found to be independent of the value of $A$, which has been set
to be $A=1,2$ and $3$.

\begin{figure}
  \begin{center}
     \includegraphics[width=3.35in,angle=0]{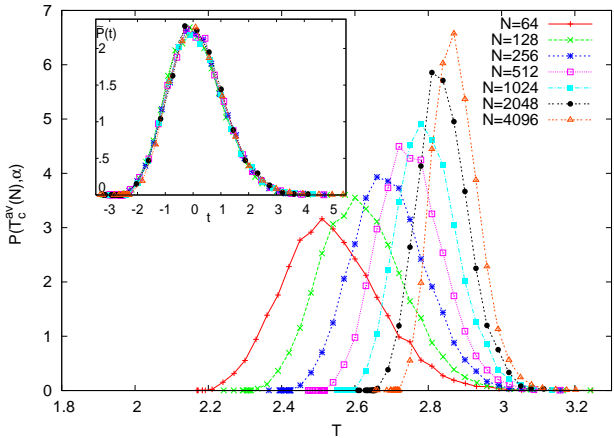}
   \end{center}
   \caption{(Color online) Distribution of the finite-size transition temperatures for different
sizes of the BA network. Inset: scaling collapse of the data in terms of $t=(T_c(N)-T^{av}_c(N))/\Delta T_c(N)$, using the scaling form in Eqs.(\ref{T_c_L}) and (\ref{DT_c_L}) with $\nu'=3.8(2)$, and $\tilde{\nu}'=5.6(2)$.
}
   \label{ptc}
 \end{figure}

\begin{figure}
  \begin{center}
     \includegraphics[width=3.35in,angle=0]{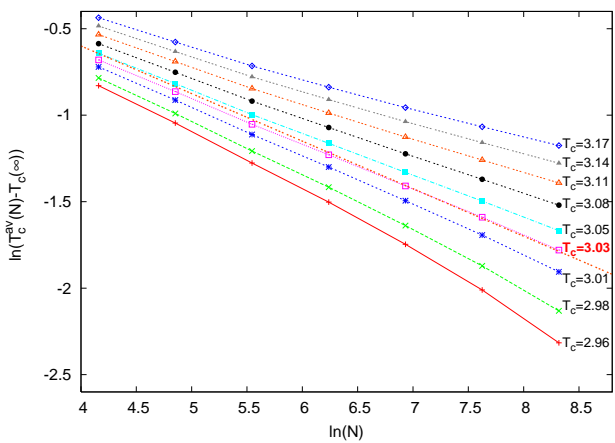}
   \end{center}
   \caption{(Color online) Shift of the average finite-size transition temperatures,
$T^{av}_c(N)-T_c(\infty)$, vs. $N$ in a log-log scale plotted for different values of $T_c(\infty)$.
The lines connecting the points at the same $T_c(\infty)$ are guide for the eye.
At the true transition point the asymptotic behavior is linear which is indicated by
a dotted straight line.
}
   \label{tc}
 \end{figure}

The distribution of the finite-size critical temperatures calculated with $x_2=0.69$ and $A=3.$
are shown in Fig.\ref{ptc} for different sizes of the network. One can observe a shift
of the position of the maxima as well as a shrinking of the width of the distribution
with increasing size of the network. The
shift of the average value, $T^{av}_c(N)$, is asymptotically given by:
\be
T^{av}_c(N)-T_c(\infty) \sim N^{-1/\tilde{\nu}'}\;,
\label{T_c_L}
\ee
whereas the width, characterized by the mean standard deviation, $\Delta T_c(N)$, scales with
another exponent, $\nu'$, as:
\be
\Delta T_c(N) \sim N^{-1/{\nu'}}\;.
\label{DT_c_L}
\ee
Using Eq.(\ref{T_c_L}) from a three-point fit we have obtained $\tilde{\nu}'=3.8(2)$ and $T_c(\infty)=3.03(2)$.
We have determined the position of the transition point in the infinite system, $T_c(\infty)$, in
another way by plotting the difference $T^{av}_c(N)-T_c(\infty)$ vs. $N$ in a log-log scale
for different values of $T_c(\infty)$, see Fig.\ref{tc}. At the true transition point according to
Eq.(\ref{T_c_L}) there is an asymptotic linear dependence, which is indeed the case around $T_c(\infty)=3.03(2)$
and the slope of the line is compatible with $1/\tilde{\nu}'=.27(1)$.

For the width exponent, $\nu'$, we obtained from Eq.(\ref{DT_c_L}) with two-point fit the estimate: $\nu'=5.6(2)$.
With these parameters the data in Fig.\ref{ptc} can be collapsed to a master curve
as shown in the inset of Fig.\ref{ptc}. This master curve looks 
not symmetric, at least for the finite sizes used
in the present calculation, and can be well fitted by a modified Gumbel distribution,
$G_{\omega}(-y)=\omega^{\omega}/\Gamma(\omega)(\exp(-y-e^{-y}))^{\omega}$,
with a parameter $\omega=4.2$. We note that the same type of fitting curve has already been used in Ref.\cite{mg05}.
For another values of the initial parameter, $A=1$ and $2$ the estimates of
the critical exponents as well as the position of the transition point are found to be stable and stand in the
range indicated by the error bars.

The equations in Eqs.(\ref{T_c_L}) and (\ref{DT_c_L}) are generalizations of the relations
obtained in regular $d$-dimensional lattices~\cite{wd95,ah96,psz97,wd98,ahw98} in which
$N$ is replaced by $L^d$, $L$ being the linear size of the system and therefore instead
of $\nu'$ and $\tilde{\nu}'$ we have $\nu=\nu'/d$ and $\tilde{\nu}=\tilde{\nu}'/d$, respectively.
Generally at a random fixed point the two characteristic exponents are equal and satisfy
the relation\cite{ccfs} $\nu'=\tilde{\nu}' \ge 2$. This has indeed been observed for
the $2d$\cite{long2d} and $3d$\cite{mai05,mai06}
random bond Potts models for large $q$ at disorder induced critical points.
On the other hand if the transition stays first-order there are two distinct exponents~\cite{fisher,mg05}
$\tilde{\nu}'=1$ and $\nu'=2$.

Interestingly our results on the distribution of the finite-size transition temperatures in
networks are different of those found in regular lattices. Here the transition is of second order
but still there are two distinct critical exponents, which are completely different of that at
a disordered first-order transition. For our system $\nu'>\tilde{\nu}'$, which means that
disorder fluctuations in the critical point are dominant over deterministic shift of the
transition point. Similar trend is observed about the finite-size transition parameters in the
random transverse-field Ising model\cite{ilrm07}, the critical behavior of which is controlled by
an infinite disorder fixed point. In this respect the RBPM in scale-free networks can be
considered as a new realization of an infinite disorder fixed point.

\subsection{Size of the critical cluster}
\label{cluster}

\begin{figure}
  \begin{center}
     \includegraphics[width=3.35in,angle=0]{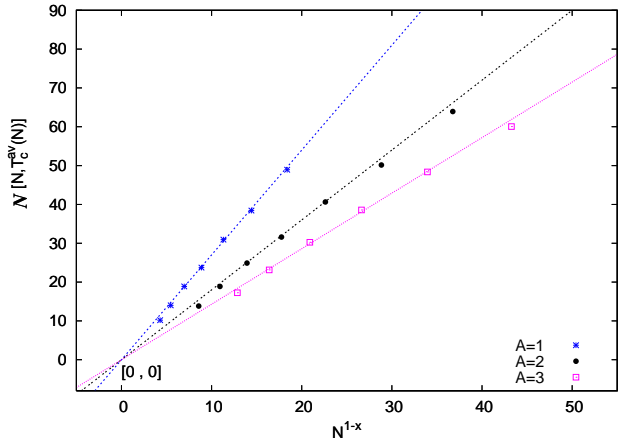}
   \end{center}
   \caption{(Color online) Size dependence of the critical cluster at the
average finite-size critical temperature as a function of $N^{1-x}$ with $x=.65$.
The date points for different
initial parameters, $A$, are well described by straight lines, which are guides
to the eye.
}
   \label{clust}
 \end{figure}

Having the distribution of the finite-size transition temperatures
we have calculated the size of the largest cluster at $T^{av}_c(N)$,
which is expected to scale as ${\cal N}[N,T^{av}_c(N)] \sim N^{1-x}$. Then from
two-point fit we have obtained an estimate for the magnetization exponent: $x=.66(1)$.
We have also plotted ${\cal N}[N,T^{av}_c(N)]$ vs. $N^{1-x}$ in Fig.\ref{clust}
for different initial parameters, $A$. Here we have obtained an asymptotic linear
dependence with an exponent, $x=.65(1)$, which agrees with the previous value within
the error of the calculation.

\section{Discussion in terms of optimal cooperation}
\label{sec_disc}

In this paper we have studied the properties of the Potts model for large value of $q$
on scale-free evolving complex networks, such as the BA network, both for homogeneous and random
ferromagnetic couplings. This problem is equivalent to an optimal cooperation problem
in which the agents try to optimize the total sum of the benefits coming from pair
cooperations (represented by the Potts couplings) and the total sum of the support
which is the same for each cooperating projects (given by the temperature of the
Potts model). The homogeneous problem is shown exactly to have two distinct states:
either all the agents cooperate with each other or there is no cooperation between any agents.
There is a strongly first-order phase transition: by increasing the support the agents
stop cooperating at a critical value.

In the random problem, in which the benefits are random and depend on the pairs
of the cooprerating agents, the structure of the optimal set depends on the value
of the support. Typically the
agents are of two kinds: a fraction of $m$ belongs to a large cooperating cluster whereas
the others are isolated, representing one man's projects. With increasing support
more and more agents are split off the cluster, thus its size, as well as $m$ is decreasing,
but the cluster keeps its scale-free
topology. For a critical value of the support $m$ goes to zero continuously and the
corresponding singularity is characterized by non-trivial critical exponents. This transition,
as shown by the numerically calculated critical exponents for the BA network, belongs to a
new universality class. One interesting feature of it is that the distribution of the finite-size
transition points is characterized by two distinct exponents and the width of the distribution is
dominated over the shift of the average transition point, which is characteristic at an
infinite disorder fixed point\cite{ilrm07}.

We thank for useful discussions with C. Monthus and for previous cooperation on the subject
with M.-T. Mercaldo.
This work has been supported by the National Office of Research and Technology under
Grant No. ASEP1111, by the
Hungarian National Research Fund under grant No OTKA TO48721, K62588, MO45596 and M36803.
M.K. thanks the Minist\`ere Fran\c{c}ais des
Affaires \'Etrang\`eres for a research grant.


\begin{thebibliography}{99}

\bibitem{AB01}
  R. Albert and A.L. Barab\'asi, Rev. Mod. Phys. {\bf 74}, 47 (2002).

\bibitem{DM01}
  S.N. Dorogovtsev and J.F.F. Mendes, Adv. Phys. {\bf 51}, 1079 (2002).
  
\bibitem{DM03} S.N. Dorogovtsev and J.F.F. Mendes, {\it Evolution of
  Networks: From Biological Nets to the Internet and WWW} (Oxford
  University Press, Oxford, 2003)

\bibitem{N03}
M.E.J. Newman, SIAM Review 45, 167-256 (2003).

\bibitem{bollobas}
  B. Bollob\'as, {\it Random Graphs} (Academic Press, London, 1985).

\bibitem{WS98}
  D.J. Watts and S.H. Strogatz, Nature (London) {\bf 393}, 440 (1998).

\bibitem{BA99}
  A.L. Barab\'asi and R. Albert, Science {\bf 286}, 509 (1999).

\bibitem{epidemic}
  R. Pastor-Satorras and A. Vespignani, Phys. Rev. Lett. {\bf 86}, 3200 (2001);
  Phys. Rev. E {\bf 63}, 066117 (2001).

\bibitem{newman}
 M.E.J. Newman, Phys. Rev. E {\bf 66}, 016128 (2002).

\bibitem{dezso}
Z. Dezs\H o, A.-L. Barab\'asi, Phys. Rev. E {\bf 65}, 055103 (2002).

\bibitem{percolation}
  D.S. Callaway, M.E.J. Newman, S.H. Strogatz, and D.J. Watts, Phys. Rev. Lett.
  {\bf 85}, 5468 (2000); R. Cohen, D. ben-Avraham, and S. Havlin,
  Phys. Rev. E {\bf 66}, 036113 (2002).

\bibitem{diffusion}
J.-D. Noh, and H. Rieger, Phys. Rev. Lett. {\bf 92}, 118701 (2004).

\bibitem{kji06}
M. Karsai, R. Juh\'asz, and F. Igl\'oi, Phys. Rev. E{\bf 73}, 036116 (2006).

\bibitem{zrp}
J.-D. Noh, Phys. Rev. E{\bf 72}, 056123 (2005).

\bibitem{stauffer}
  A. Aleksiejuk, J.A. Holyst, and D. Stauffer, Physica A {\bf 310}, 260  (2002).

\bibitem{ising}
  M. Leone, A. Vazquez, A. Vespignani, and R. Zecchina, Eur. Phys. J. B {\bf 28}, 191 (2002);
  S.N. Dorogovtsev, A.V. Goltsev, and J.F.F. Mendes, Phys. Rev. E {\bf 66}, 016104 (2002);
  G. Bianconi, Phys. Lett. A {\bf 303}, 166 (2002).

\bibitem{sajat}
 F. Igl\'oi and L. Turban, Phys. Rev. E{\bf 66}, 036140 (2002).

\bibitem{potts}
 S.~N. Dorogovtsev, A.~V Goltsev, and J.F.F. Mendes,  Eur. Phys. J. B {\bf 38}, 177 (2004).

\bibitem{joseph} C.V. Giuraniuc, J.P.L. Hatchett, J.O. Indekeu, M.
  Leone, I. Perez Castillo, B. Van Schaeybroeck, and C. Vanderzande,
Phys. Rev. Lett. {\bf 95}, 098701 (2005); Phys. Rev. E {\bf 74}, 036108 (2006).

\bibitem{sumour}
M.A. Sumour, A.H. El-Astal, F.W.S. Lima, M.M. Shabat, H.M. Khalil, e-print cond-mat/0612189.

\bibitem{aizenmanwehr}
        M. Aizenman and J. Wehr, Phys. Rev. Lett. {\bf 62}, 2503
        (1989); errata {\bf 64}, 1311 (1990).

\bibitem{uzelac}
 K. Uzelac, A. Hasmy, and R. Jullien, Phys. Rev. Lett. {\bf 74}, 422 (1995).

\bibitem{pottssite}
H.G. Ballesteros, L.A. Fern\'andez, V. Mart\`in-Mayor, A. Mu\~noz Sudupe,
G. Parisi, and J.J. Ruiz-Lorenzo, Phys. Rev. B {\bf 61}, 3215 (2000).

\bibitem{pottsbond}
        C. Chatelain, B. Berche, W. Janke, and P.-E. Berche, Phys. Rev. E{\bf 64}, 036120 (2001);
        W. Janke, P.-E. Berche, C. Chatelain, and B. Berche, Nuclear Physics B {\bf 719} 275 (2005).

\bibitem{mai05}
   M.-T. Mercaldo, J-Ch. Angl\'es d'Auriac, and F. Igl\'oi, Europhys. Lett. {\bf 70}, 733 (2005).

\bibitem{mai06}
   M.-T. Mercaldo, J-Ch. Angl\'es d'Auriac, and F. Igl\'oi, Phys. Rev. E{\bf 73}, 026126 (2006).

\bibitem{aips02}
        J.-Ch. Angl\`es d'Auriac {\it et al.}, J. Phys. A{\bf 35}, 6973 (2002);
        J.-Ch. Angl\`es d'Auriac, in {\it New Optimization Algorithms in Physics},
        edt. A.~K. Hartmann and H. Rieger (Wiley-VCH, Berlin 2004).

\bibitem{JRI01}
  R. Juh\'asz, H. Rieger, and F. Igl\'oi, Phys. Rev. E{\bf 64}, 056122 (2001).

\bibitem{kasteleyn}
        P.W. Kasteleyn and C.M. Fortuin, J. Phys. Soc. Jpn. {\bf 46} (suppl.),
        11 (1969).

\bibitem{Wu}
  F.Y. Wu, Rev. Mod. Phys. {\bf 54}, 235 (1982).

\bibitem{long2d}
        M.-T. Mercaldo, J-Ch. Angl\'es d'Auriac, and F. Igl\'oi, Phys. Rev. E {\bf 69}, 056112 (2004).

\bibitem{gls81}
        M. Gr\"otschel, L. Lov\'asz, A. Schrijver, Combinatorica 1 169-197 (1981).

\bibitem{ai03}
  J.-Ch. Angl\`es d'Auriac and F. Igl\'oi, Phys. Rev. Lett. {\bf90}, 190601 (2003).

\bibitem{ER}
P. Erd\H os and A. R\'enyi, Publ. Math. Debrecen {\bf 6}, 290 (1959);
Publ. Math. Inst. Hung. Acad. Sci. {\bf 5}, 17 (1960).

\bibitem{mg05}
  C. Monthus, and T. Garel, Eur. Phys. J. B {\bf 48}, 393 (2005).

\bibitem{wd95}
S. Wiseman and E. Domany, Phys Rev E {\bf 52}, 3469 (1995).

\bibitem{ah96}
A. Aharony, A.B. Harris, Phys Rev Lett {\bf 77}, 3700 (1996).

\bibitem{psz97} F. P\'azm\'andi, R.T. Scalettar and G.T. Zim\'anyi,
Phys. Rev. Lett. {\bf 79}, 5130 (1997).

\bibitem{wd98}
S. Wiseman and E. Domany, Phys. Rev. Lett. {\bf 81}, 22 (1998) ; Phys Rev E {\bf 58}, 2938 (1998).

\bibitem{ahw98}
A. Aharony, A.B. Harris and S. Wiseman,
Phys. Rev. Lett. {\bf 81}, 252 (1998).

\bibitem{ccfs}
        J. T. Chayes {\it et al.},
        Phys. Rev. Lett. {\bf 57}, 299 (1986).

\bibitem{fisher}
        D.S. Fisher, Phys. Rev. B {\bf 51}, 6411 (1995).

\bibitem{ilrm07}
	F. Igl\'oi, Y.-C. Lin, H. Rieger, and C. Monthus, (unpublished).

\end{thebibliography}
\end{document}